\RequirePackage{fix-cm}
\documentclass[natbib, smallcondensed]{svjour3}     % onecolumn (ditto)
\smartqed  % flush right qed marks, e.g. at end of proof
\usepackage{graphicx}
\usepackage{subcaption}
\usepackage{float}
\usepackage{rotating}
\usepackage{lscape}
\usepackage{capt-of}
\usepackage{multirow}
\begin{document}

\title{An Empirical Study of Compression-friendly Community Detection Methods}


\author{Muhammad Irfan Yousuf, Izza Anwer, Muhammad Abid}

\authorrunning{M. I. Yousuf et al.}

\institute{M. I. Yousuf \at
              Department of Computer Science (New Campus) \\
              University of Engineering and Technology, Lahore, Pakistan\\
              Tel: +92-42-37951901\\
              \email{irfan.yousuf@uet.edu.pk}\\          
              I. Anwer \at
              Department of Transportation Engineering and Management\\
              University of Engineering and Technology, Lahore, Pakistan\\
              M. Abid \at
              Department of Computer and Information Sciences\\
              Pakistan Institute of Engineering and Applied Sciences, Islamabad, Pakistan            
}

\date{Received: date / Accepted: date}

\maketitle

\begin{abstract}
Real-world graphs are massive in size and we need a huge amount of space to store them. Graph compression allows us to compress a graph so that we need a lesser number of bits per link to store it. Of many techniques to compress a graph, a typical approach is to find clique-like caveman or traditional communities in a graph and encode those cliques to compress the graph. On the other side, an alternative approach is to consider graphs as a collection of hubs connecting spokes and exploit it to arrange the nodes such that the resulting adjacency matrix of the graph can be compressed more efficiently. 
We perform an empirical comparison of these two approaches and show that both methods can yield good results under favorable conditions. We perform our experiments on ten real-world graphs and define two cost functions to present our findings.  
\end{abstract}

\section{Introduction}
A graph illustrates interconnections among entities. Mathematically, graphs are structures that represent pairwise relations between objects. A graph has two main components: nodes; which represent objects, edges; that represent relationships between nodes. A graph can have millions or even billions of nodes or edges in it and storage of such massive graphs require a huge amount of memory. To store big graphs, data compression is one of the possible techniques to reduce the number of bits required to represent a graph. Data compression can dramatically decrease the amount of storage a file takes up. However, the question is how can we compress graphs efficiently?

Many of the data reduction methods exploit the redundancy in data to compress it. A closer look at the well-known representations of graphs e.g., adjacency lists or adjacency matrix shows that we can exploit the redundancies in these representations to lower the number of bits required to store a graph \citep{Adler_2001,Boldi_2004,Raghavan_2003,Randall_2002,Suel_2001}. Of the possible solutions, finding communities in a graph and then ordering the nodes such that the nodes with the same neighborhood are ordered in close proximity has shown good results \citep{Buehrer_2008,Lim_2014}. We can say that the problems of compressing a graph and finding communities in it are closely related. If we can find good communities, we can compress the graph efficiently. 

In this paper, we investigate the impact of community-based node ordering on graph compression. While research related to community detection dates back to the 70s in mathematical sociology and circuit design \citep{Algo_1972,Struct_1972}, Newman’s and Girvan’s work on modularity in complex systems just a couple of decades ago revitalized the field of community detection, making it one of the main pillars of network science research \citep{Newman_2004,Newman_2006}. Taking into account its importance, it is not surprising that many community detection methods have been developed \citep{Newman_2002,Clauset_2004,Blondel_2008,Rosvall_2008,Raghavan_2007,newman_2006_2,Reichardt_2006, Pons_2005} using tools and techniques from variegated disciplines such as statistical physics, biology, applied mathematics, computer science, and sociology. Most recently, the authors \citep{Lim_2014} proposed a new approach to finding communities designed specifically for producing a compression-friendly ordering. The traditional community detection methods \citep{Raghavan_2007,Blondel_2008,Clauset_2004,newman_2006_2,Rosvall_2008} focus on finding homogeneous regions in the graph so that nodes inside a region are tightly connected to each other than to nodes in other regions. However, the authors \citep{Lim_2014} go beyond the traditional methods and exploit the power-law degree distribution of real-world graphs to find compression-friendly communities. 

In this work, we compare the compression ratios achieved by deploying some of the traditional state-of-the-art community detection methods to find communities \citep{Newman_2002,Clauset_2004,Blondel_2008,Rosvall_2008,Raghavan_2007,newman_2006_2,Reichardt_2006, Pons_2005} versus the recent approach to find compression-friendly communities \citep{Lim_2014}.  Our purpose is twofold:

\begin{itemize}
	\item Comparing the different traditional community detection methods in terms of finding good communities for graph compression.
	\item Comparing traditional approaches with the recent approach to community detection for better compression ratios.
\end{itemize}

The rest of the paper is organized as follows. In section 2, we provide some basic definitions to understand graphs and community detection methods. In section 3, we define graph compression and propose a naive node ordering scheme. In section 4, we present our findings. In section 5, we discuss the related work and conclude the paper in section 6.   

\begin{figure}[t]
	\centering
	\includegraphics[width = 120mm, height=60mm]{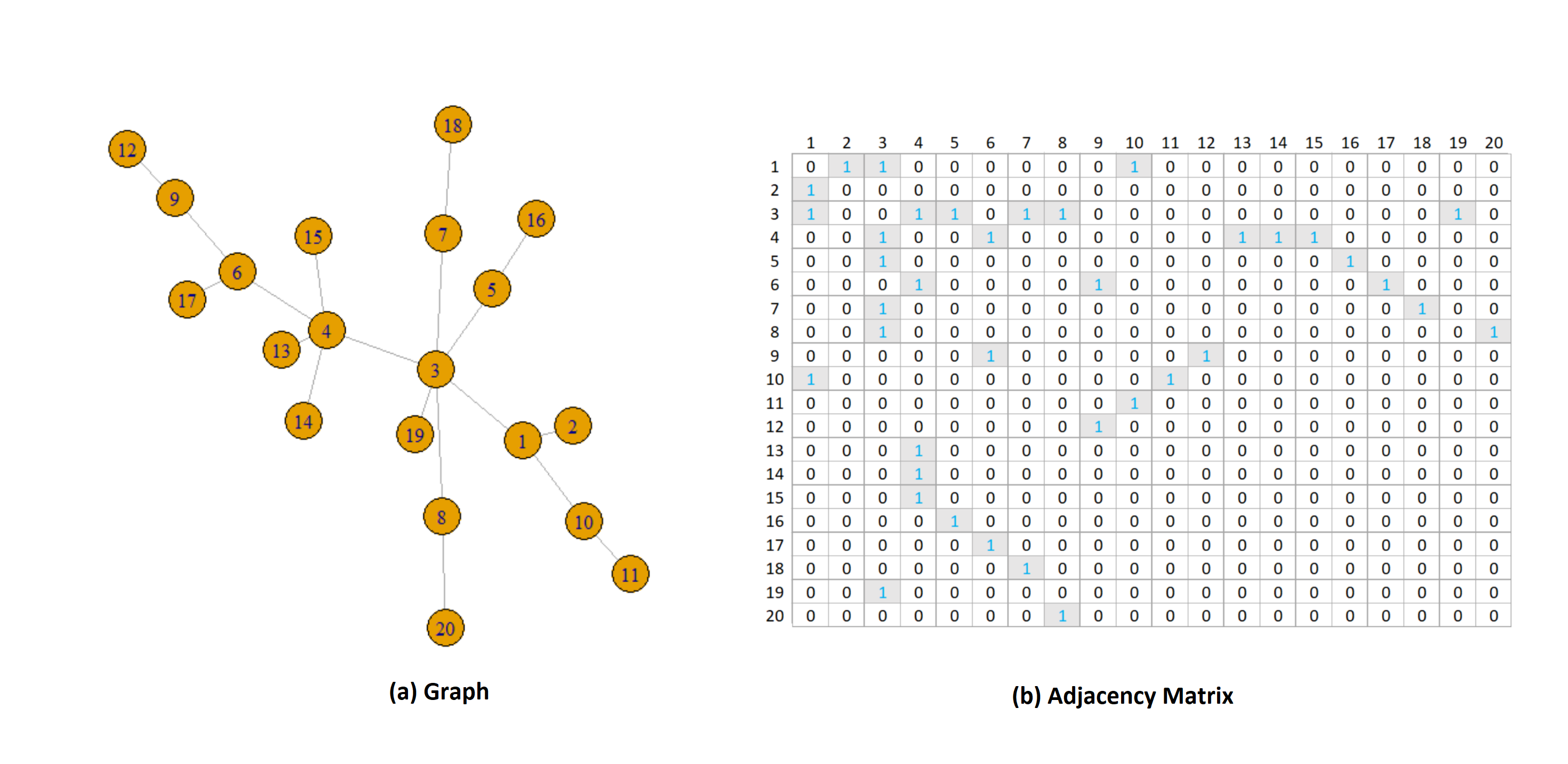}
	\caption{ (a) A small graph (b) its adjacency matrix}
	\label{fig_1}
\end{figure}

\section{Basic Definitions and Preliminaries}
In this section, we provide some basic definitions and preliminaries.
\subsection{Graphs and Their Representation}
Graphs are data structures and offer a way of expressing relationships between objects. Graphs are used to represent networks, e.g., social networks, biological networks, collaboration networks, etc. A graph consists of nodes (or vertices) and edges (or links) between them to show relationships. Mathematically, a graph is written as $G(V, E)$ where $V$ is the set of nodes and $E$ is the set of edges. There are different ways to represent a graph but in this paper, we will consider the adjacency matrix representation of a graph.  Further, we consider only undirected graphs in this work. A small graph and its corresponding adjacency matrix are shown in Figure~\ref{fig_1}.

\subsection{Community Detection Methods}
A community in a graph forms a group of nodes such that these nodes have more connections with one another than with the nodes in the rest of the graph. Community structures are very common in real networks. Social networks include communities based on common location, interests, occupation, etc. In the last two decades, many methods have been proposed to detect non-overlapping communities in graphs. We select the following five traditional methods for this work. In addition, we compare these traditional methods with a recent approach, SlashBurn \citep{Lim_2014}, specifically designed to find compression-friendly communities in a graph. 

\begin{figure}[t]
	\centering
	\includegraphics[width = 120mm, height=60mm]{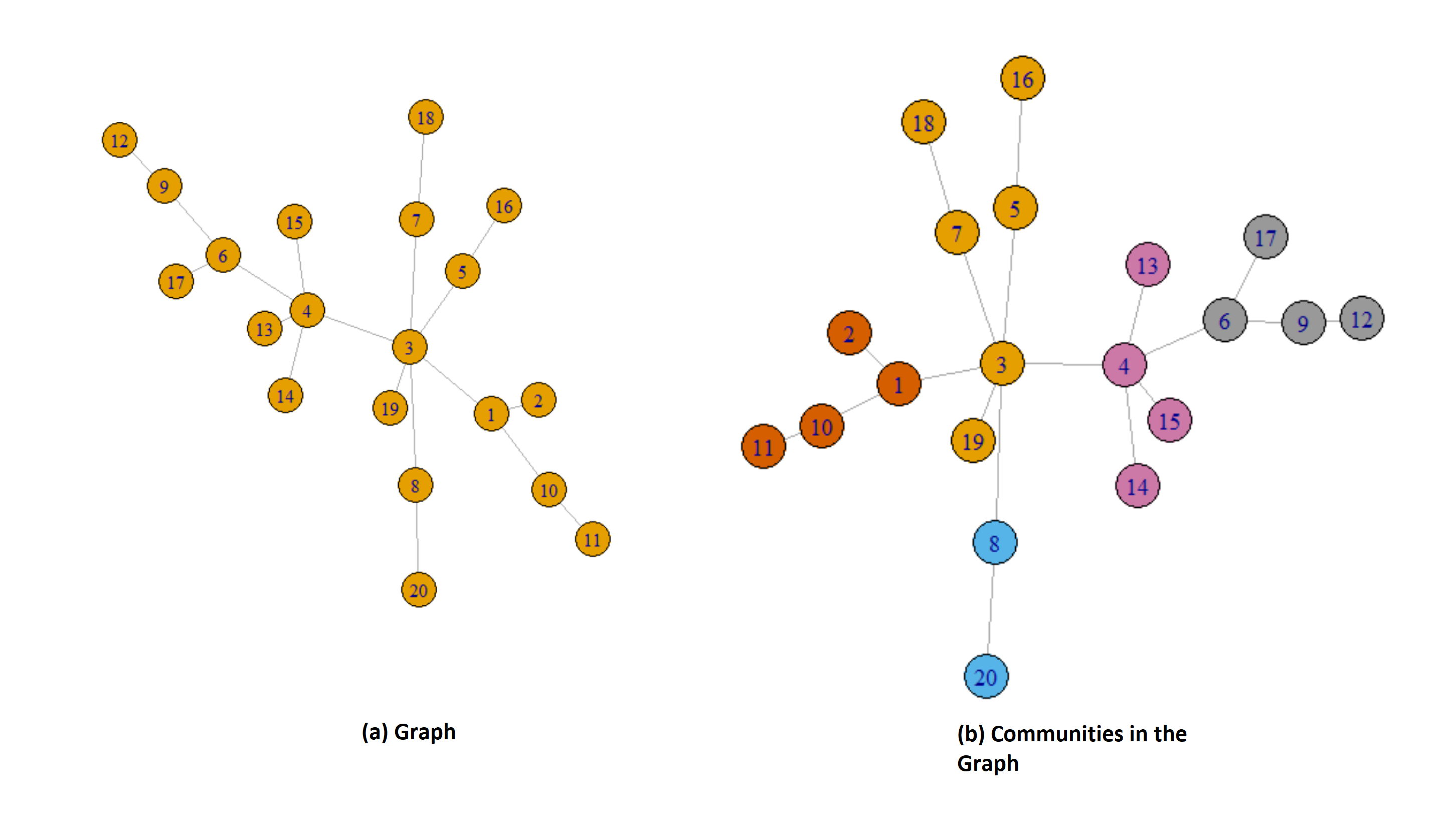}
	\caption{ (a) A small graph (b) Communities found in it with MultiLevel method.}
	\label{fig_2}
\end{figure}

In Figure~\ref{fig_2}, we show a small graph and the communities found in it using the MultiLevel \citep{Blondel_2008} method of finding non-overlapping communities. We find five communities in the graph and show the nodes with a different color belonging to a community. 

\subsubsection{Label Propagation}
This algorithm \citep{Raghavan_2007} assumes that each node in the network is assigned to the same community as the majority of its neighbors. At startup, it gives a distinct label to each node in the network. As the algorithm progresses, each node takes the label of the majority of its neighbors. The above step stops once each node has the same label as the majority of its neighbors. 

\subsubsection{MultiLevel}
MultiLevel \citep{Blondel_2008} is a greedy approach to optimize modularity. This method first assigns a different community to each node of the network, then a node is moved to the community of one of its neighbors to improve modularity. The above step is repeated for all nodes until no further improvement can be achieved. Then each community is considered as a single node on its own and the second step is repeated until there is only a single node left or when the modularity cannot be increased in a single step. 

\subsubsection{Fast Greedy}
Fast Greedy \citep{Clauset_2004} is also a greedy community detection method that optimizes the modularity score. It starts with a non-clustered initial assignment where each node belongs to a one-node community. Next, it computes the expected improvement of modularity for each pair of communities and merges them into one if that pair gives the maximum improvement of modularity. The above procedure is repeated until we cannot improve modularity by merging communities.

\subsubsection{Leading Eigenvector}
Leading Eigenvectors \citep{newman_2006_2} method optimizes the modularity by using the eigenvalues and eigenvectors of the modularity matrix. First, the leading eigenvector of the modularity matrix is calculated, and then the graph is divided into two parts such that modularity improvement is maximized based on the leading eigenvector. Repeatedly, the modularity contribution is calculated at each step in the subdivision of a network. It stops once the positive contribution to modularity is not possible. 

\subsubsection{Infomap}
Infomap \citep{Rosvall_2008} employs random walks to analyze the information flow through a network. Initially, it encodes the network into modules in a way that maximizes the amount of information about the original network. Then it sends the signal to a decoder through a channel with limited capacity. The decoder tries to construct a set of possible candidates for the original graph after decoding the message. The smaller the number of candidates, the more information about the original network has been transferred.

\subsubsection{SlashBurn}
In principle, SlashBurn \citep{Lim_2014} is a node-reordering algorithm specifically developed for graph compression. It goes beyond the traditional ‘caveman communities’ and finds communities by removing high degree nodes from the graph. The notion behind the algorithm is that removing the highest centrality nodes from the graph produces many small disconnected components that could be considered as a ‘community’. First, it removes high centrality nodes from the graph. Next, it reorders nodes such that high-degree nodes are assigned the lowest IDs, nodes from disconnected components are assigned the highest IDs, and nodes from the giant connected component are assigned the middle-range of IDs. In the next iteration, the process is repeated on the giant connected component. The process stops when the size of the giant component is smaller than the number of nodes to be deleted in the next iteration. 

\section{Graph Compression}
The main objective of a graph compression algorithm is to reduce the number of bits required to store the graph. The recent research \citep{Chierichetti_2009, Boldi_2011} shows that the ordering of nodes plays an important role in compressing graphs. Given a graph, we reorder the nodes such that a smaller number of bits are required to store the graph than the number of bits required for the unordered graph. Formally, we define our problem as: 
\begin{quote}
	\textit{Given a graph with adjacency matrix A, we find an ordering of nodes such that the storage cost function is minimized.}
\end{quote}
In general, the solution is to arrange the entries in the adjacency matrix so that we could group them in such a way that a group has (ideally) only 1s or 0s in it. For this purpose, we reorder the nodes such that the nodes belonging to a cluster (or community) get consecutive positions in the ordering. With such a compression-friendly ordering, we can compress the adjacency matrix by omitting the groups that have only 0s in it and compressing only those groups (or blocks) that have 1s in it.

In Figure~\ref{fig_3}, we show the adjacency matrix of a small graph (see Fig.~\ref{fig_1}) with (a) random ordering of nodes and (b) compression-friendly node ordering. We apply 2 by 2 blocks to cover all the nonzero elements inside the matrix. we find that the right matrix requires a smaller number of blocks (24 blocks) than the left matrix (34 blocks). It is clear that the matrix on the right has a lesser number of non-empty blocks than the left matrix and therefore, we need a lesser number of bits to store the right matrix than the left one although both matrices are of the same graph.

\subsection{Node Ordering}
\begin{figure}[t]
	\centering
	\includegraphics[width = 120mm, height=60mm]{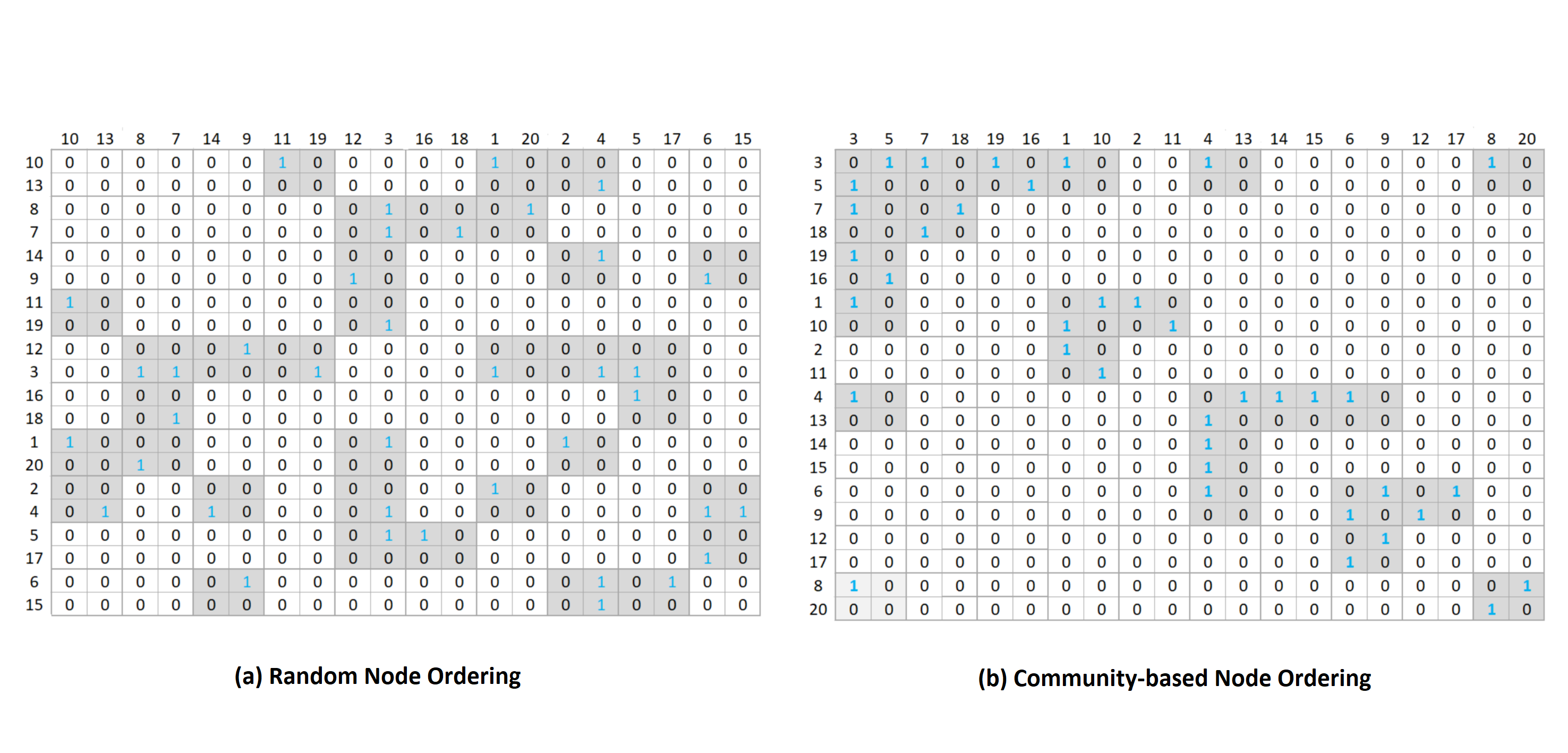}
	\caption{Adjacency matrix of a graph when the nodes are arranged  (a) Randomly (b) Based on the size of communities and degree of nodes.}
	\label{fig_3}
\end{figure}

In the problem statement, we mentioned that we need to find a node ordering that reduces the storage cost function. We compare the five traditional community detection methods, i.e., Label Propagation, MultiLevel, Fast Greedy, Leading Eigenvectors, and Infomap with the latest node ordering and community detection method, i.e., SlashBurn. For the traditional community detection methods, we propose a naive node ordering as follows. 
\begin{itemize}
	\item In the first step, we find communities in a graph using one of the community detection methods.
	\item In the second step, we order the nodes in descending order of community sizes they belong to, i.e., the nodes belonging to the largest community are at the top of the ordered list whereas the nodes of the smallest community sink to the bottom of the ordered list.
	\item In the last step, we order the nodes within a community in the descending order of their degrees.
\end{itemize}
To sum up, we order the nodes in descending order of community sizes and within a community, we order in descending order of degrees of nodes. The adjacency matrix of Figure~\ref{fig_3}(b) follows this node ordering.

To compare with the SlashBurn method, we follow the same node ordering as described in the original paper \citep{Lim_2014}. The SlashBurn method removes hub nodes to create spokes and a giant connected component (GCC). The next iteration starts on the GCC. The method works iteratively until we reach a threshold. The method exploits the hubs to define alternative communities different from the traditional communities. Although this method achieves good compression ratios, there are two points to consider; 1) The graph shattering process to remove the hubs is computationally expensive. 2) The value of $k$ in the k-hubset, the set of nodes with top k highest centrality scores, and the value of $b$, block width, do not have optimal values that will suit different types of real-world graphs.  

\subsection{Cost Functions}
The main purpose of compressing a graph is to reduce the storage cost function. If we are given the adjacency matrix $A$ of a graph with compression-friendly node ordering, the cost $cost(A)$ to store this matrix should be minimum. In other words, we will need a lesser number of bits per link to store the graph. We define the following two cost functions as described in \citep{Lim_2014}. 

\subsubsection{Number of Non-empty Blocks}
Let $b$ be the width of a square block to cover all the nonzero elements in a matrix $A$. If we use $b$ by $b$ blocks to cover all the 1s in the adjacency matrix $A$, then the first cost function $cost_1(A)$ is defined as the number of nonempty blocks in $A$.

\begin{equation}
	cost_1(A) = number\; of\; nonempty\; b\; by\; b\; blocks
\end{equation}
 where $b$ is the width of a block. It is obvious that a compression-friendly node ordering should give a smaller number of non-empty blocks than random node ordering. For example, the right matrix in Figure~\ref{fig_3} has 24 non-empty blocks compared to the 34 blocks of the left matrix. 

\subsubsection{Number of Bits per Link}
The second cost function $cost_2(A)$ uses the number of bits per link (or edge) required to encode the adjacency matrix using a block-wise encoding. We define this cost function assuming a compression method achieving the information-theoretic lower bound \citep{Chakra_2004}.

\begin{equation}
	cost_2(A) = |B|. 2log\frac{n}{b} + \sum_{\beta \in B}b^2.H(\frac{z(\beta)}{b^2})
\end{equation}

where $n$ is the number of nodes, $B$ is the set of non-empty blocks of width $b$, $z(\beta)$ is the number of 1s within block $\beta$ and the function $H(.)$ is the binary Shannon entropy function defined as:

\begin{equation}
	H(p) = plog\frac{1}{p} + (1 - p)log\frac{1}{1-p}
\end{equation}
The first term in Equation (2) gives the number of bits to encode meta-information of blocks, i.e., row and column  IDs of blocks whereas the second term calculates the number of bits to store 1s in the nonempty blocks. For example, we need 248 bits to store the right matrix of Figure~\ref{fig_3} while the left matrix will consume 340 bits on storage. 

\section{Experimental Setup and Results}
We apply the traditional community detection methods to find communities in a graph and then order the nodes as described above to get a compression-friendly adjacency matrix. Similarly, we also order the nodes as described in the SlashBurn method. We then find both the cost functions to compare these methods. The graphs used in our experiments are described in Table~\ref{Tab_ds}. The graphs are publicly available \citep{Kdata,Ndata}.

\begin{table}
	\centering
	\caption {Real-world graphs used in the experiments} 
	\begin{tabular}{|l|l|l|p{6cm}|}
		\hline
		DataSet & Nodes & Edges & Description \\ \hline
		Facebook & 63,731 & 817,089 & A friendship network extracted from facebook consisting of people with edges representing  friendship ties. \\ 
		BlogCatalog & 88,784 & 2,093,193 & BlogCatalog is a social blog directory. The dataset contains all links among users. \\ 
		LiveMocha & 104,103 & 2,193,081 & LiveMocha is the largest language learning community in the world.  The dataset contains all links among users. \\ 
		Academia & 200,188 & 1,022,484 & Academia.edu is a platform for academics  to share research papers. The dataset is a social network of its users. \\ 
		GooglePlus & 211,336 & 1,141,861 & GooglePlus is a social networking service offered by Google. The network is a friendship network among its users. \\ 
		TwitterHiggs & 456,635 & 12,508,466 & The dataset is used to study the spreading process on Twitter before, during and after the announcement of the discovery of Higgs Boson in 2012. \\ 
		Delicious & 536,108 & 1,365,959 & Delicious is a social bookmarking service for storing, sharing and discovering web bookmarks. \\
		Last.fm & 1,191,805 & 4,519,330 & This is social network of friendships of last.fm music website. The nodes are users and edges represent friendships among  them. \\ 
		YouTube & 1,134,890 & 2,987,624 & Friendship network of video sharing website YouTube. \\ 
		Hyves & 1,402,673 & 2,777,919 & This is a social network of a Dutch online community. Nodes are users and Egdes are their friendships. \\ \hline
	\end{tabular}
	\label{Tab_ds}
\end{table}

\begin{figure}[t]
	\centering
	\includegraphics[width = 120mm, height=80mm]{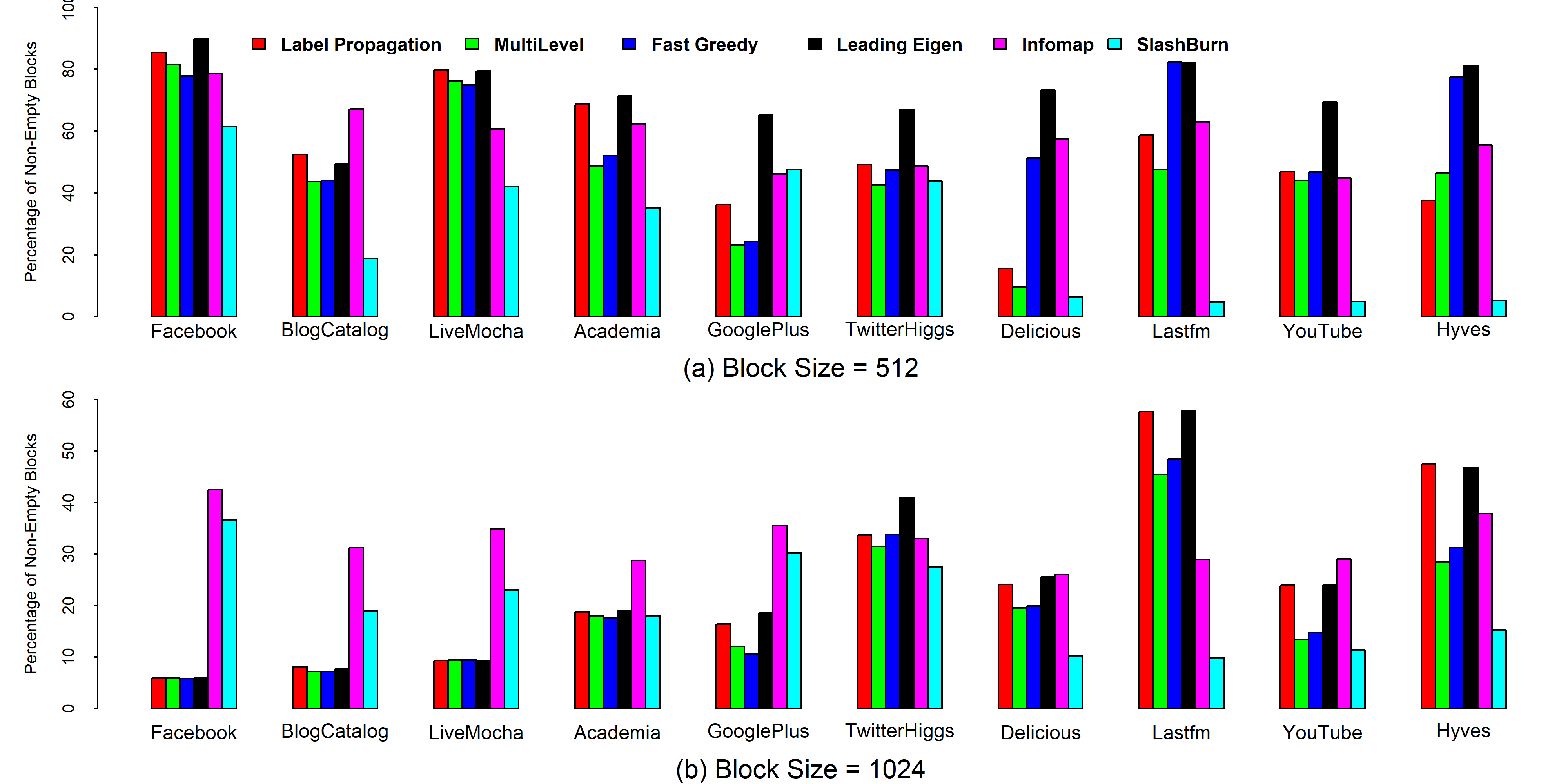}
	\caption{$cost_1(A)$:Number of non-empty blocks for block widths (a) b = 512 and (b) b = 1024}
	\label{fig_4}
\end{figure}

\subsection{Counting Non-empty Blocks}
In this experiment, we count the number of non-empty blocks in the adjacency matrix of each graph for all the methods. We set the block width $b=512$ and $b=1024$ so that we also see the effect of block size on this cost function. We present the results in Figure~\ref{fig_4} in the form of bar graphs where the height of a bar shows the percentage of non-empty blocks. In Figure~\ref{fig_4}(a), we see that $cost_1(A)$ is minimum in the SlashBurn method in all the datasets except GooglePlus. In the last four datasets it has a very small value while in other datasets, it gives relatively higher values. On the other side, in Figure~\ref{fig_4}(b), $cost_1(A)$ is minimum in only four datasets in the SlashBurn method. Although SlashBurn was specially designed to produce a compression-friendly node ordering, it seems that it does not minimize the first cost function $cost_1(A)$ for any value of $b$.  This experiment also shows that the traditional community detection methods combined with a naive node ordering can yield a minimum number of non-empty blocks in the adjacency matrix of a graph.

\begin{table}
	\centering
	\caption {$cost_2(A)$: The number of Bits per Link achieved by all the methods when $b$ = 512. \textbf{Boldface} values show the best results.}
	\begin{tabular}{|p{1.6cm}|p{1.6cm}|p{1.2cm}|p{1.2cm}|p{1.4cm}|p{1.2cm}|p{1.2cm}|}
		\hline
		DataSet & Label Propagation & Multi Level & Fast Greedy & Leading Eigen & InfoMap & SlashBurn \\ \hline
		Facebook & 10.66 & \textbf{9.99} & 10.39 & 10.58 & 11.67 & 10.36 \\ 
		BlogCatalog & 8.37 & 8.27 & 8.35 & 8.25 & 8.26 & \textbf{6.68} \\ 
		LiveMocha & 11.76 & 11.42 & 11.52 & 11.61 & 10.66 & \textbf{9.37} \\ 
		Academia & 14.42 & 13.24 & 13.97 & 15.29 & 15.01 & \textbf{13.06} \\ 
		GooglePlus & \textbf{9.14} & 9.24 & 9.85 & 13.14 & 12.34 & 11.12 \\ 
		TwitterHiggs & 11.08 & 10.59 & 11.36 & 12.33 & \textbf{9.14} & 10.47 \\ 
		Delicious & 14.49 & 14.09 & 14.36 & 15.76 & 18.34 & \textbf{12.77} \\ 
		Last.fm & 18.61 & 17.13 & 17.83 & 18.70 & 20.84 & \textbf{13.09} \\ 
		YouTube & 13.67 & 12.99 & 13.40 & 15.31 & 17.84 & \textbf{12.73} \\ 
		Hyves & 18.29 & 16.44 & 17.19 & 19.09 & 15.47 & \textbf{14.53} \\ \hline
		Average & 13.05 & 12.34 & 12.82 & 14.01 & 13.96 & \textbf{11.42} \\ \hline
	\end{tabular}
      \label{Tab_512}
\end{table}

\begin{table}
	\centering
	\caption {$cost_2(A)$: The number of Bits per Link achieved by all the methods when $b$ = 1024. \textbf{Boldface} values show the best results.}
	\begin{tabular}{|p{1.6cm}|p{1.6cm}|p{1.2cm}|p{1.2cm}|p{1.4cm}|p{1.2cm}|p{1.2cm}|}
		\hline
		DataSet & Label Propagation & Multi Level & Fast Greedy & Leading Eigen & Infomap & SlashBurn \\ \hline
		Facebook & 9.94 & \textbf{9.56} & 9.65 & 9.86 & 10.01 & 10.35 \\ 
		BlogCatalog & \textbf{6.83} & 7.54 & 7.18 & 6.95 & 7.44 & 6.88 \\ 
		LiveMocha & 9.43 & 9.62 & 9.62 & 9.53 & 9.68 & \textbf{9.38} \\ 
		Academia & 12.90 & \textbf{12.18} & 12.62 & 13.51 & 12.71 & 12.91 \\ 
		GooglePlus & \textbf{8.69} & 8.92 & 8.83 & 11.05 & 8.82 & 10.81 \\ 
		TwitterHiggs & 9.48 & \textbf{9.38} & 9.72 & 10.45 & 9.41 & 10.39 \\ 
		Delicious & 13.08 & 12.56 & \textbf{12.50} & 14.34 & 14.63 & 12.82 \\ 
		Last.fm & 13.92 & \textbf{12.75} & 13.24 & 13.94 & 16.95 & 12.88 \\ 
		YouTube & 13.22 & 12.67 & 12.92 & 14.49 & 15.78 & \textbf{12.51} \\ 
		Hyves & 15.62 & \textbf{13.46} & 13.97 & 15.29 & 17.47 & 13.90 \\ \hline
		Average & 11.31 & \textbf{10.86} & 11.02 & 11.94 & 12.29 & 11.28 \\ \hline
	\end{tabular}
     \label{Tab_1024}
\end{table}

\subsection{Calculating Bits per Link}
In this experiment, we calculate the number of bits per link required to store a graph as per Equation (2) under two block widths $b=512$ and $b=1024$. We present the results in Tables~\ref{Tab_512} and~\ref{Tab_1024}. We see that when $b=512$, on average, SlashBurn achieves the minimum number of bits per link to store a graph. However, when $b=1024$ MultiLevel yields the best results on average. These results show that SlashBurn does not perform consistently and the number of bits achieved by it depends on the value of $b$. In other words, we need to check the value of $b$ if we want to achieve good results with the SlashBurn. The same is true for other methods. It seems that going beyond caveman or traditional communities does not always offer benefits rather a good node ordering scheme combined with a traditional community detection method and an optimal value of $b$ can yield good compression ratios.

\section{Related Work}
There has been a lot of works on community detection and graph compression. A community detection method aims at finding homogeneous regions in the graph so that inter-region edges between different regions are minimized \citep{Raghavan_2007,Blondel_2008,Clauset_2004,newman_2006_2,Rosvall_2008}. In this context, network modularity is a closely related topic as it measures the strength of division of a graph into groups or communities. This is the reason these two topics usually go hand in hand \citep{Newman_2006}.
Community detection in graphs or networks has applications in many areas of research including finance \citep{finance1,finance2}, medical \citep{medical1,medical2}, criminology \citep{crim1,crim2}, and transportation \citep{transport1, transport2}, to name a few. Graph compression has also been an active research topic and several techniques have been developed to compress large graphs. The authors \citep{Boldi_2004} proposed graph compression techniques to encode large web graphs whereas \citep{Chierichetti_2009} extended the work to include social networks. These studies apply reference encoding to compress adjacency lists of big graphs. The authors \citep{Maneth_2016} compress graphs by grammar while the study \citep{Grabowski-2011} merges adjacency lists of a graph to store it efficiently. 
\section{Conclusion}
In this work, we propose a naive node ordering scheme and show that we can achieve good compression ratios when the nodes belonging to a traditional community are arranged according to this naive node ordering. We consider five traditional community detection methods and a recent approach that exploits hubs and spokes in a graph to define alternative communities. We define two cost functions to measure the goodness of a compression-friendly community detection scheme. We perform experiments on ten real-world graphs and show that clique-like caveman communities are also compression-friendly when combined with a good node ordering. Both approaches can produce good results under suitable conditions.

\bibliographystyle{spbasic}      % basic style, author-year citations
\bibliography{ComCompression}   % name your BibTeX data base

\end{document}